\documentclass[%
reprint,
superscriptaddress,
showpacs,preprintnumbers,
amsmath,amssymb,
aps,
prb,
notitlepage,
citeautoscript,
]{revtex4-1}

\usepackage{graphicx}
\usepackage[export]{adjustbox}
\usepackage{dcolumn}
\usepackage{bm}
\usepackage{amsmath} 
\usepackage{wasysym}
\usepackage[utf8]{inputenc}
\usepackage{color}
\usepackage{lipsum}
\usepackage{epstopdf}

\renewcommand{\vec}[1]{
\mathbf{#1}
}

\graphicspath{{graphics/}}

\begin{document}
\newlength\fheight \newlength\fwidth 


\title{Magnetic edge states and magnetotransport in graphene antidot barriers}

\author{M. R. Thomsen}
\affiliation{Department of Physics and Nanotechnology, Aalborg University, DK-9220 Aalborg \O st, Denmark}
\affiliation{Center for Nanostructured Graphene (CNG), DK-9220 Aalborg \O st, Denmark}
\author{S. R. Power}
\affiliation{Center for Nanostructured Graphene (CNG), DK-9220 Aalborg \O st, Denmark}
\affiliation{Center for Nanostructured Graphene (CNG), DTU Nanotech, Department of Micro- and Nanotechnology, Technical University of Denmark, DK-2800 Kongens Lyngby, Denmark}
\author{A.-P. Jauho}
\affiliation{Center for Nanostructured Graphene (CNG), DTU Nanotech, Department of Micro- and Nanotechnology, Technical University of Denmark, DK-2800 Kongens Lyngby, Denmark}
\author{T. G. Pedersen}
\affiliation{Department of Physics and Nanotechnology, Aalborg University, DK-9220 Aalborg \O st, Denmark}
\affiliation{Center for Nanostructured Graphene (CNG), DK-9220 Aalborg \O st, Denmark}

\begin{abstract}
Magnetic fields are often used for characterizing transport in nanoscale materials. Recent magnetotransport experiments have demonstrated that ballistic transport is possible in graphene antidot lattices (GALs). 
These experiments have inspired the present theoretical study of GALs in a perpendicular magnetic field. We calculate magnetotransport through graphene antidot barriers (GABs), which are finite rows of antidots arranged periodically in a pristine graphene sheet, using a tight-binding model and the Landauer-B\"uttiker formula. We show that GABs behave as ideal Dirac mass barriers for antidots smaller than the magnetic length, and demonstrate the presence of magnetic edge states, which are localized states on the periphery of the antidots due to successive reflections on the antidot edge in the presence of a magnetic field. We show that these states are robust against variations in lattice configuration and antidot edge chirality. Moreover, we calculate the transmittance of disordered GABs and find that magnetic edge states survive a moderate degree of disorder. Due to the long phase-coherence length in graphene and the robustness of these states, we expect magnetic edge states to be observable in experiments as well.
\end{abstract}

\maketitle

\section{Introduction}
Graphene antidot lattices (GALs), which are periodic perforations in a graphene sheet, may open a band gap in the otherwise semi-metallic material \cite{pedersen2008graphene,petersen2010clar,eroms2009weak,kim2010fabrication,oberhuber2013weak,brun2014electronic}. An advantage of GALs is that the size of the band gap can be tuned by geometrical factors. Recent magnetotransport experiments have demonstrated that ballistic transport is possible in GALs \cite{sandner2015ballistic,yagi2015ballistic}, which gives rise to interesting phenomena such as magnetoresistance oscillations due to cyclotron orbits that are commensurate with the antidot lattice. Ballistic transport in pristine graphene has been demonstrated several times and even at room temperature \cite{wang2013one,taychatanapat2013electrically,novoselov2007room,dean2011multicomponent,du2009fractional,antti2016ballistic}, but ballistic transport in GALs has previously been hindered by defects introduced by top-down fabrication of the antidots. The recent demonstration \cite{sandner2015ballistic,yagi2015ballistic} of ballistic transport in GALs was achieved by minimizing interaction with the substrate by using hexagonal boron nitride (hBN) substrates and by reducing edge roughness by encapsulating the graphene flake in hBN before etching the antidot lattice \cite{sandner2015ballistic}. 


Previous theoretical studies on nanostructured graphene in magnetic fields have primarily focused on the density of states and optical properties \cite{pedersen2013hofstadter,islamouglu2012hofstadter,zhang2008tuning,pedersen2012dirac}. The density of states of a structure under a magnetic field reveals a self-similar structure known as Hofstadter's butterfly \cite{hofstadter1976energy}. In particular, Hofstadter butterflies of GALs have revealed band gap quenching induced by perpendicular magnetic fields \cite{pedersen2013hofstadter}. Transport calculations have yet to reveal if band gap quenching also gives rise to quenching of the transport gap. 
Using the Dirac approximation, perforations in a graphene sheet are modelled as local mass terms rather than potentials\cite{brun2014electronic}. Within this description, it has been demonstrated that a single graphene antidot supports localized edge states in the presence of magnetic fields \cite{pedersen2012dirac}. Conceptually, one may think of these as edge states due to repeated reflections of electrons on the antidot edge provided the radius of the cyclotron motions is small compared to the antidot radius. We will refer to these as "magnetic edge states", not to be confused with spin-polarized edge states, such as those observed on extended zigzag edges\cite{trolle2013large}. Hence, by such states we simply mean states that are localized near an antidot due to the magnetic field. 

Magnetic edge states occur when the electron wave interferes constructively with itself in a pinned orbit around the antidot, which gives rise to Aharonov-Bohm-type oscillations. In conventional semiconductors, such as GaAs, Aharonov-Bohm oscillations due to antidots in two-dimensional electron gases have been studied theoretically \cite{takagaki1994aharonov,bogachek1995edge,ishizaka1997detailed} and bserved experimentally\cite{weiss1991electron,nihey1993aharonov,schuster1994phase}. Additionally, a theoretical study predicts the presence of Aharonov-Bohm-type oscillations in graphene nanorings \cite{farghadan2013magnetic}. We likewise predict magnetic edge states to be present in GALs and due to the long phase-coherence length in graphene, we expect these to be observable in experiments as well. 
Cyclotron orbits were recently imaged in pristine graphene using cooled scanning probe microscopy \cite{bhandari2015imaging,morikawa2015imaging}. It would be remarkable if this technique could be used for direct observation of magnetic edge states in graphene antidots.


In the present work, we study the transport properties of graphene antidot barriers (GABs), i.e., finite rows of antidots in an otherwise pristine graphene sheet, in the presence of perpendicular magnetic fields. In our transport calculations, we use the Landauer-B\"uttiker formalism with a tight-binding model, which is widely used for calculating the quantum transport in nanoscale devices \cite{xu2008magnetic,power2014electronic,thomsen2014dirac,pedersen2012graphene,thomsen2015spin,datta1995electronic,markussen2006electronic,pedersen2012transport,gunst2011thermoelectric}. The magnetic field is included in the Hamiltonian by a Peierls substitution. The calculations utilize the recursive Green's function (RGF) method, which greatly reduces the calculation time, while retaining accuracy. 
Furthermore, we compare the tight-binding results to both an ideal Dirac mass barrier and a gapped graphene model. 
We find that Dirac mass barriers provide a good description of the transport gap for GABs with small antidots provided the magnetic field is not too strong. Furthermore, we find evidence of magnetic edge states on the antidots and demonstrate simple scaling of these, allowing predictions for larger systems. Finally, we calculate the transmittance of disordered GABs and compare this to the corresponding transmittance in ordered GABs.

\section{Theory and methods}
\subsection{Tight-binding model}
In this section, we will use the RGF method with a tight-binding model in order to calculate transmittance of electrons through GABs in a magnetic field. The barrier regions are periodic perpendicular ($y$-direction) to the transport direction ($x$-direction). We also calculate the density of states (DOS) of fully periodic GALs and compare these to the transmittance of GABs.

\begin{figure}[htb]
\centering
\def\svgwidth{1\columnwidth}
\includegraphics[width=1\columnwidth]{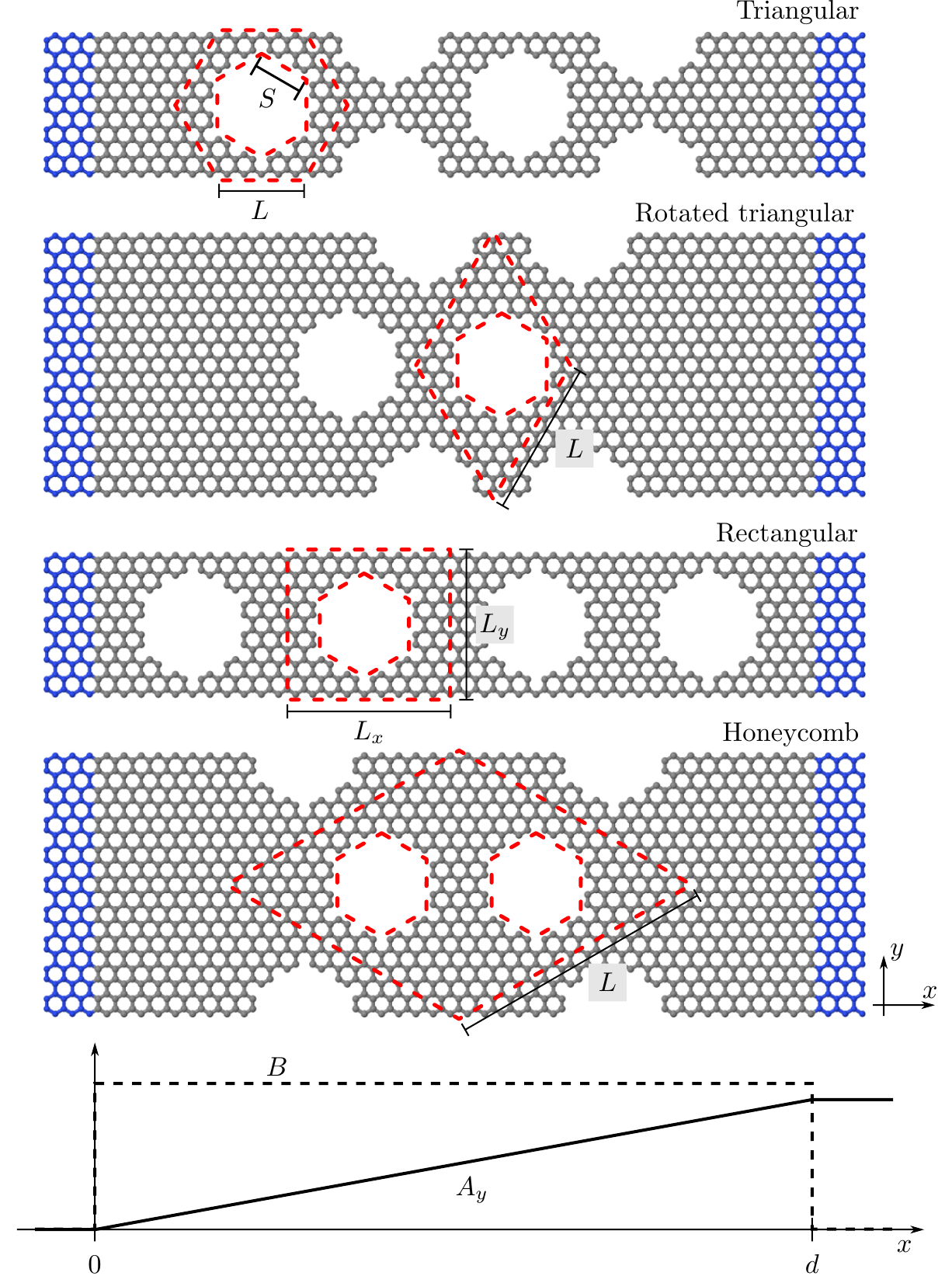}
\caption{GAB unit cells used in transport calculations and corresponding vector potential and magnetic field. The unit cells shown here all have four rows of antidots in the transport direction, the same antidot size and similar neck widths. The gray and blue atoms represent the system and semi-infinite leads, respectively. The dashed red lines outline the corresponding GAL unit cells.}
\label{fig:vector_potential}
\end{figure}

In the nearest-neighbor orthogonal tight-binding model, the Hamiltonian can be written as

\begin{equation}
\hat H = \sum_{i<j} t_{ij} \hat c_i^\dagger \hat c_j + H.c.,
\end{equation}
where the hopping parameter $t_{ij}$ is taken as $-\gamma$ for nearest neighbors and vanishing otherwise. 
The magnetic field is included by performing the Peierls substitution $t_{ij}\rightarrow t_{ij}e^{i\phi_{ij}}$, where $\phi_{ij}=(e/\hbar)\int_{\vec{r}_i}^{\vec{r}_j} \vec{A}\cdot d\vec{l}$ is the Peierls phase, $\vec A$ is the vector potential, and $\vec{r}_i$ is the position of atom $i$. 
The magnetic field in the leads is taken to zero, which means the vector potential in the Landau gauge is given by 
\begin{equation}
	\vec A(\vec r) = \hat{\vec{y}}B\bar{x},\quad \bar{x}=\left\{ \begin{array}{ll}
0,\,&x<0\\
x,\,& 0\leq x \leq d\\
d,\,&x>d
\end{array}\right. ,
\label{eq:landau_gauge}
\end{equation}
where $d$ is the width of the barrier, see Fig.~\ref{fig:vector_potential}. Note that the vector potential can not be set to zero in the $x>d$ region, as this would imply an infinite magnetic field at the $x=d$ interface. 
In this gauge, the Peierls phase becomes

\begin{equation}
	\phi_{ij} = \frac{eB}{2\hbar}(y_j-y_i)(\bar{x}_i+\bar{x}_j).
\label{eq:landau_gauge_potential}
\end{equation}

We present calculations for triangular, rotated triangular, rectangular and honeycomb GALs in the notation of Ref.~\onlinecite{petersen2010clar}. 
We will use hexagonal antidots with armchair edges and denote the antidot lattices by $\{L,S\}$, where $L$ and $S$ are the side lengths, in units of the graphene lattice constant $a=0.246$~Å, of the GAL unit cell and the antidot, respectively, see Fig.~\ref{fig:vector_potential}. For rectangular lattices, we use $L_x$ and $L_y$ to denote the side lengths in the $x$- and $y$-directions, respectively. In our calculations, we chose $L_y\approx L_x = L$ in order for the unit cell to be approximately square. Unless stated otherwise, calculations are made on triangular GABs and assume periodic boundary conditions along the $y$-direction. Calculations on GALs also assume periodic boundary conditions along the $x$-direction and the results are $k$-averaged in the periodic directions. The number of $k$-points in each direction is taken as the odd integer closest to $400/L$. 

We also perform calculations on a gapped graphene model, where instead of introducing antidots, a band gap is opened by using a staggered sublattice potential of $\Delta$ on one sublattice and $-\Delta$ on the other, opening a band gap of $E_g=2\Delta$ \cite{pedersen2009optical}. The advantage of this method compared to using the actual antidot geometry is that it is computationally much faster due to the reduced width of the unit cell in the $y$-direction.

We use the RGF method to extract properties such as transmittance and DOS. This method has the same accuracy as direct diagonalization, but is considerably faster. 
The method is outlined in Refs.~\onlinecite{svizhenko2002two,lewenkopf2013recursive} and relies on calculating certain block elements of the retarded Green's function $G=((E+i\varepsilon)I-H-\Sigma_L-\Sigma_R)^{-1}$ by slicing the system into smaller cells, which only couple to themselves and their nearest neighbors. $H$ is the Hamiltonian matrix and $\Sigma_L$ and $\Sigma_R$ are the self energies of the semi-infinite pristine graphene left and right leads, respectively. Also, $i\varepsilon$ is a small imaginary factor added to the energy. While $\varepsilon$ should, in principle, be infinitesimal we apply a finite but small value for numerical stability and in practise take $\varepsilon=\gamma10^{-4}$ in all calculations. The lead self energies are omitted when calculating the DOS of the GALs, as these are additionally periodic along the $x$-direction. The GAL unit cells are indicated by the dashed red lines in Fig.~\ref{fig:vector_potential}. The RGF algorithms require the Hamiltonian to be block tridiagonal. In case of GABs the Hamiltonian is block tridiagonal by construction, but in case of GALs it is not, due to periodicity in the $x$-direction coupling the first cell to the last ($N$th) one. In this case, the Hamiltonian can easily be made block tridiagonal by merging cells such that cells 1 and $N$ are merged, 2 and $N-1$ are merged and so forth. The result is that the diagonal blocks double in size, but the resulting matrix is block diagonal. 

For GALs, we also have to ensure periodicity of the Peierls phase, due to the additional periodicity of the system in the $x$-direction. This limits the $B$-fields that can be used in a calculation, but is remedied by creating a supercell consisting of several unit cells as was also done in Ref.~\onlinecite{pedersen2013hofstadter}. The minimal $B$-field which ensures periodicity of the Peierls phase is denoted $B_{min}$. The $B$-field is then written as $B=n B_{min}$, where $n$ is an integer. When the magnetic flux $\Phi=B\sqrt{3}a^2/2$ through a graphene unit cell equals one flux quantum $\Phi_0=h/e$, the energy spectrum is restored. Therefore, we only let the relative magnetic flux density $\Phi/\Phi_0\in [0;1]$. The $n$ at which the relative flux is unity is denoted $n_{max}$. The minimal field is summarized for the different lattice configurations in Tab.~\ref{tab:Bmin}. 
In practice, we take advantage of the fact that a given $B$-field can be obtained by several supercell sizes and then always choosing the smallest, as was done in Ref.~\onlinecite{pedersen2013hofstadter}. 

\begin{table}[htb]
\centering
\begin{tabular}{lccc}
\hline
 \textbf{Lattice configuration}  & $d$       & $B_{min}$ [$2h/\sqrt 3ea^2$]\\ \hline
 Triangular 			& $3LNa$ 	& $1/LN$	\\ 
 Rotated triangular 	& $LNa$ 	& $3/LN$ 	\\ 
 Rectangular			& $L_xNa$ 	& $3/L_xN$  \\ 
 Honeycomb 				& $3LNa$ 	& $1/LN$	\\\hline
\end{tabular}
\caption{The $B$-field is written as $B=nB_{min}$, where $B_{min}=h/(edy_{min})$ is the minimal $B$-field that satisfies periodicity of the Peierls phase, with $y_{min}=a/2\sqrt{3}$ for transport in the zigzag direction. The $n$ at which the relative flux is unity is given by $n_{max}=2h/(\sqrt{3}ea^2B_{min})$.} 
\label{tab:Bmin}
\end{table}

The local DOS on atom $i$ is proportional to the diagonal element of the Green's function, 

\begin{equation}
	L_i(E) = -\frac{1}{\pi}\text{Im}\{G_{ii}\},
\end{equation}
and the full DOS is then the sum of all local contributions, 

\begin{equation}
D(E)=\sum_i L_i(E).	
\end{equation}
The conductance of the system is given by the Landauer-B\"uttiker formula $G=\frac{2e^2}{h}T$, where $T=\text{Tr}\{\Gamma_L G^\dagger \Gamma_R G\}$ is the transmittance. Finally, the bond current between atom $i$ and $j$ at low temperature and low bias $V_a$ can be calculated as \cite{todorov2002tight,power2014electronic} 

\begin{equation}
I_{i\rightarrow j}(E)=-\frac{4e^2V_a}{\hbar}\text{Im}\{H_{ij}A^{(L)}_{ji}\},	
\end{equation}
where $A^{(L)}=G\Gamma_LG^\dagger$ is the left-lead spectral function.

\subsection{Magnetic edge states}
\begin{figure}[htb]
\centering
\def\svgwidth{1\columnwidth}
\includegraphics[width=1\columnwidth]{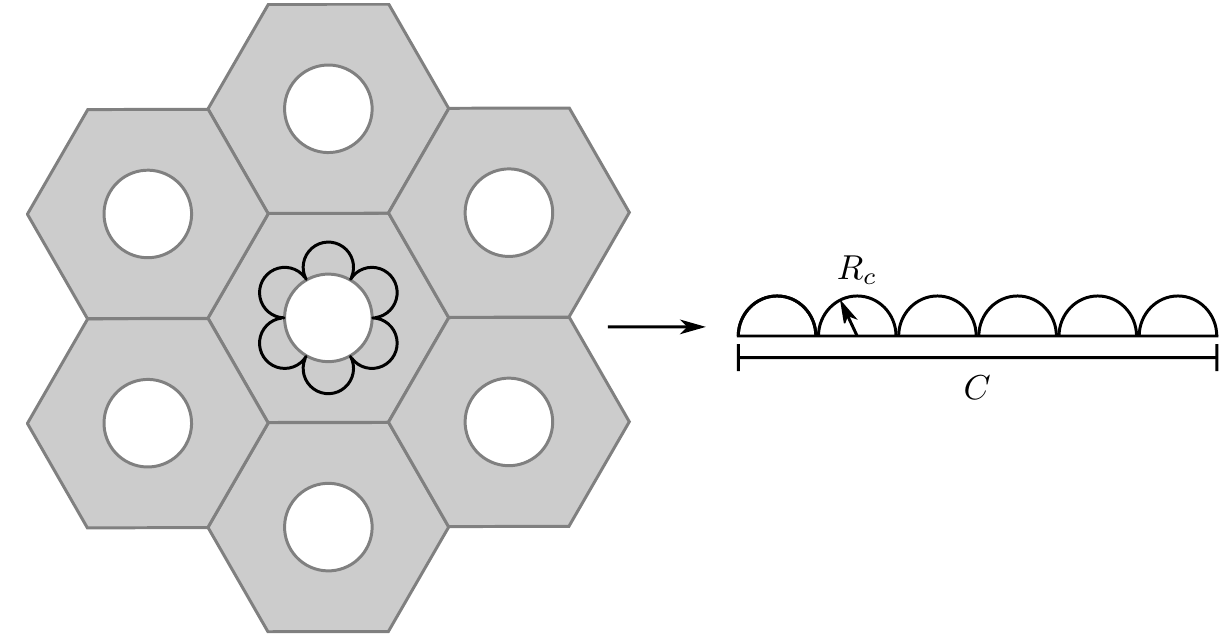}
\caption{Magnetic edge state with cyclotron radius $R_c$ for an antidot with circumference $C$.}
\label{fig:magnetic_edge_state}
\end{figure}

A prominent feature of GALs is the presence of magnetic edge states. Semi-classically, a magnetic edge state is a state which is confined to the antidot due to repeated reflections off the antidot due to the presence of an applied magnetic field as illustrated in Fig.~\ref{fig:magnetic_edge_state}. In this section, we derive an approximate condition for the occurrence of magnetic edge states. To this end, we will rely on a simple continuum (Dirac) model of gapped graphene. In this model, the energy is given by $E=\pm \sqrt{\hbar^2v_F^2k^2+\Delta^2}$, where $v_F=\sqrt{3}a\gamma/2\hbar\simeq 10^6$~m/s is the Fermi velocity.

The cyclotron radius is given by $R_c = m^*v/eB$,\cite{ashcroft1976solid} where $v$ is the speed of the electron and $m^ *$ is cyclotron effective mass (or dynamical mass), which is semi-classically given by \cite{neto2009electronic,ashcroft1976solid,smrcka1994plane}

\begin{equation}
m^* = \frac{\hbar^2}{2\pi}\left[\frac{\partial A(E)}{\partial E}\right]_{E=E_F}.
\end{equation}
Here, $A(E)$ is the area enclosed by the orbit in $k$-space and given by $A(E)=\pi k^2(E)$ for rotationally symmetric band structures. 
In the gapped graphene model, we can write $\hbar v_Fk(E)=\sqrt{E^2-\Delta^2}$, and so

\begin{equation}
A(E)= \frac{\pi (E^2-\Delta^2)}{\hbar^2 v_F^2}, \quad |E|\geq\Delta.
\end{equation}
The cyclotron effective mass is then
\begin{equation}
m^* = \frac{E}{v_F^2},\quad |E|\geq\Delta,
\end{equation}
which is exactly the same result as for pristine graphene \cite{bhandari2015imaging,neto2009electronic}. The cyclotron effective mass is thus independent of band gap, given by $E_g=2\Delta$. It does therefore not change between the pristine graphene in the leads and the antidot regions as long as the energy satisfies $|E|\geq\Delta$. The cyclotron radius is then given by 

\begin{equation}
R_c = \frac{E}{ev_F B}.
\label{eq:cyclotron_radius}
\end{equation}

In order to have a magnetic edge state, the electron must form a stationary wave on the periphery on the antidot. 
As an approximation, we analyze the case where the electron is reflected off a straight line with length equal to the circumference of the antidot $C$ and require that $2nR_c=C$, where $n$ is an integer equal to the number of reflections on a round trip, see Fig.~\ref{fig:magnetic_edge_state}. 
The $B$-fields that satisfy this requirement with $n$ reflections are then $B_n=2nE/ev_FC$. In addition, we require the electron wave function to be in phase after one orbit. The electron gains a phase on one orbit of $\phi=\int_{P}\vec{k}\cdot d\vec l=kD$, where $P$ is the path traveled by the electron and $D=n\pi R_c=\pi C/2$ is the total distance traveled. We thus require $kD=m2\pi$, where $m$ is an integer. Here, we use the approximation $\hbar v_F k=\sqrt{E^2-\Delta^2}\approx E$, which is a good approximation when $E\gg\Delta$. The energies that satisfy the phase requirement are then $E=4m\hbar v_F/C$ and we may finally write the $B$-field requirement as
\begin{equation}
	B_n=\frac{8mn\hbar}{eC^2}.
	\label{eq:magnetic_edge_state}
\end{equation}
The oscillation period of magnetoresistance caused by magnetic edge states is then given by $\Delta B = 8m\hbar/eC^2$. We see that doubling the antidot circumference, equivalent of quadrupling the area, decreases the oscillation period by a factor of four.

\section{Results}

\begin{figure}
\centering
\includegraphics[width =1\columnwidth]{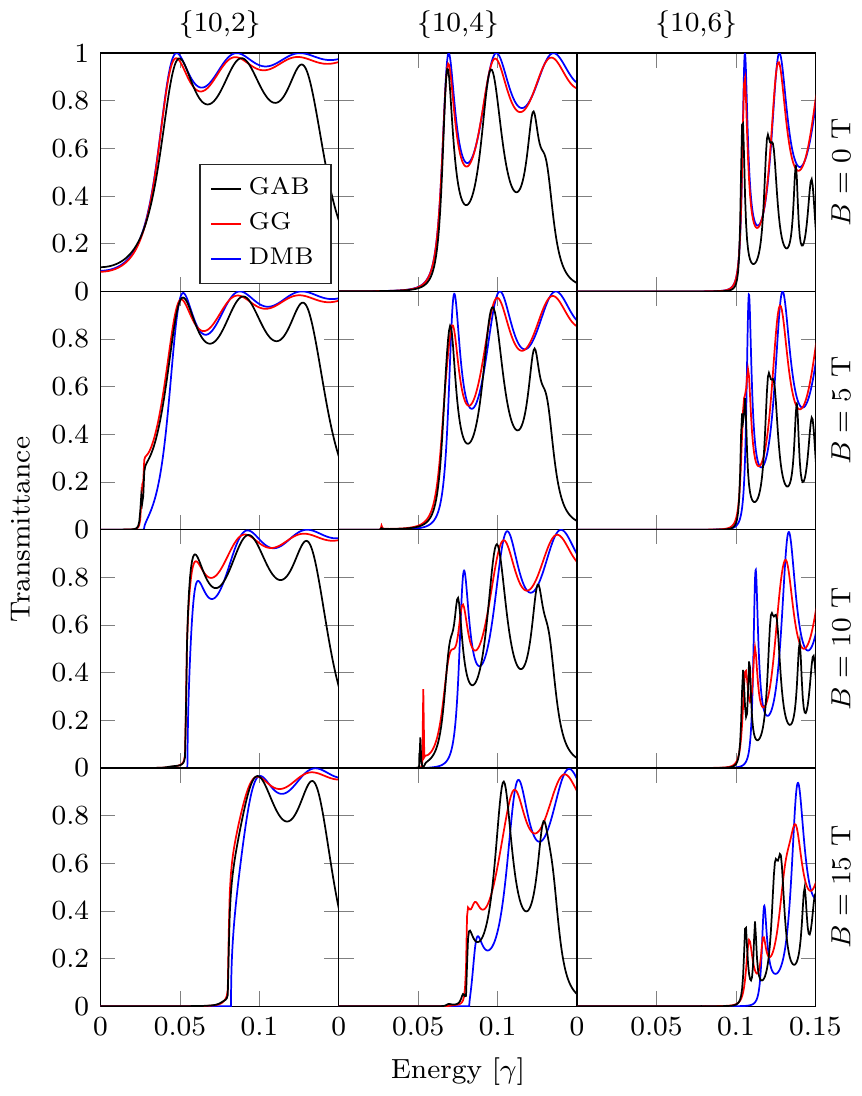}
\caption{Transmission through $\{10, S\}$ triangular GABs containing 4 rows of antidots in the transport direction, as well as gapped graphene (GG) barriers and Dirac mass barriers (DMBs) with the same length ($d=16.5$ nm) and band gaps as the GABs. All calculations were made for $k_y=0$. The TB calculations are divided by two for comparison with the single valley Dirac result.}
\label{fig:k0_trans}
\end{figure}

\begin{figure*}[htbp]
	\centering
	\includegraphics[width=0.975\textwidth]{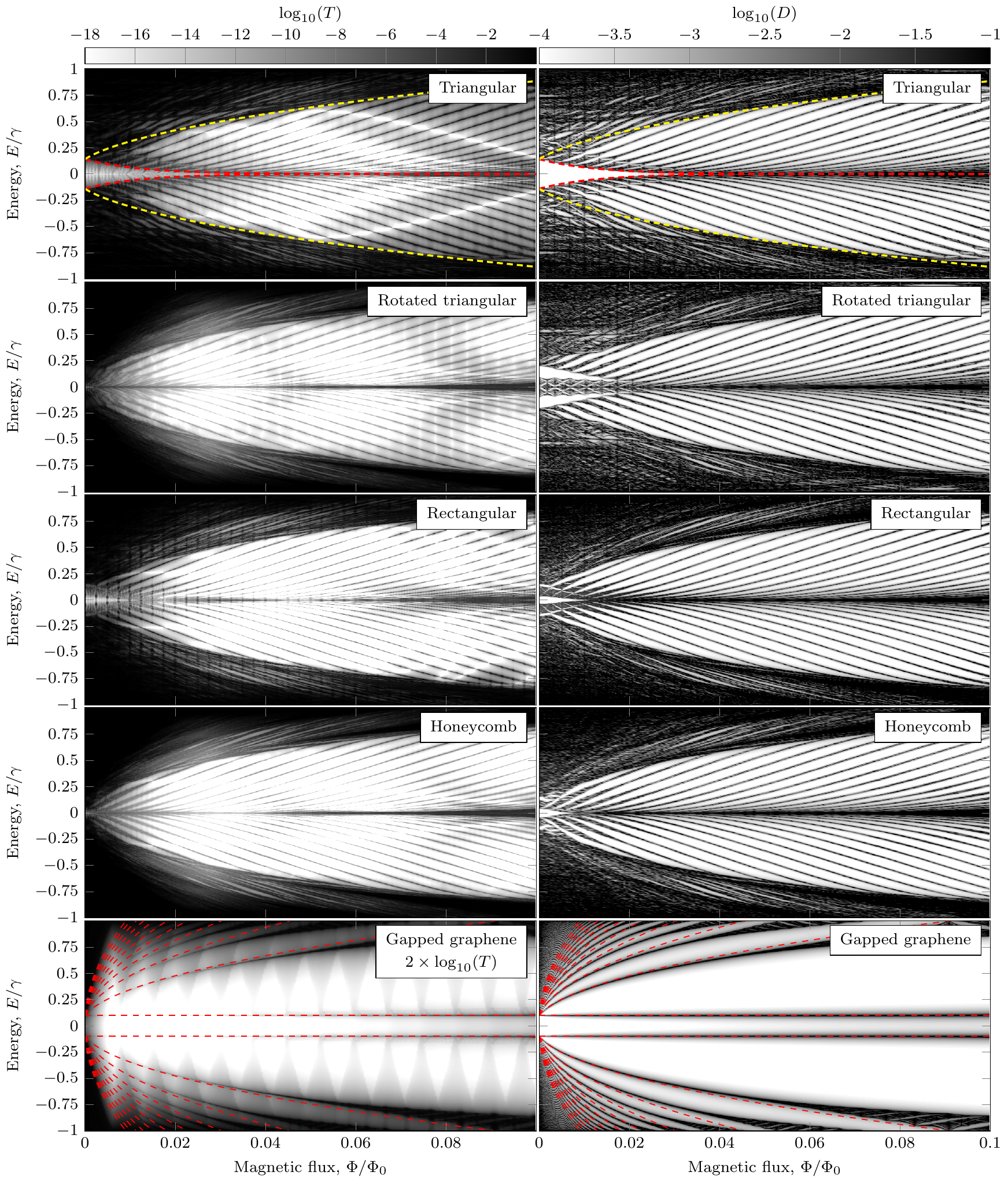}
	\caption{Comparison between transmittance (left) and DOS (right) of $\{L,6\}$ GABs in different lattice configurations. $L$ is chosen to give the systems approximately the same neck width ($\simeq1.3$ nm). For the triangular antidot lattice, this corresponds to a \{10,6\} system. The transport calculations are made with 4 rows of antidots in the transport direction. The dashed lines in the top panels outline the geometric band gap (red) and the Landau level gap (yellow). The two bottom panels show a $\Delta=0.1\gamma$ gapped graphene system. The dashed red lines in the bottom panels show the first 10 Landau levels of massive Dirac fermions $E_n = \pm \sqrt{\Delta^2 + 2v_F^2\hbar e B n}$ \cite{pedersen2013hofstadter}. For the gapped graphene model, we plot $2\times\log_{10}(T)$ due to the generally lower transmittance for this system.}
	\label{fig:transmittance_dos_comparison_armchair}
\end{figure*}
\begin{figure*}[htb]
	\centering
	\includegraphics[width=1\textwidth]{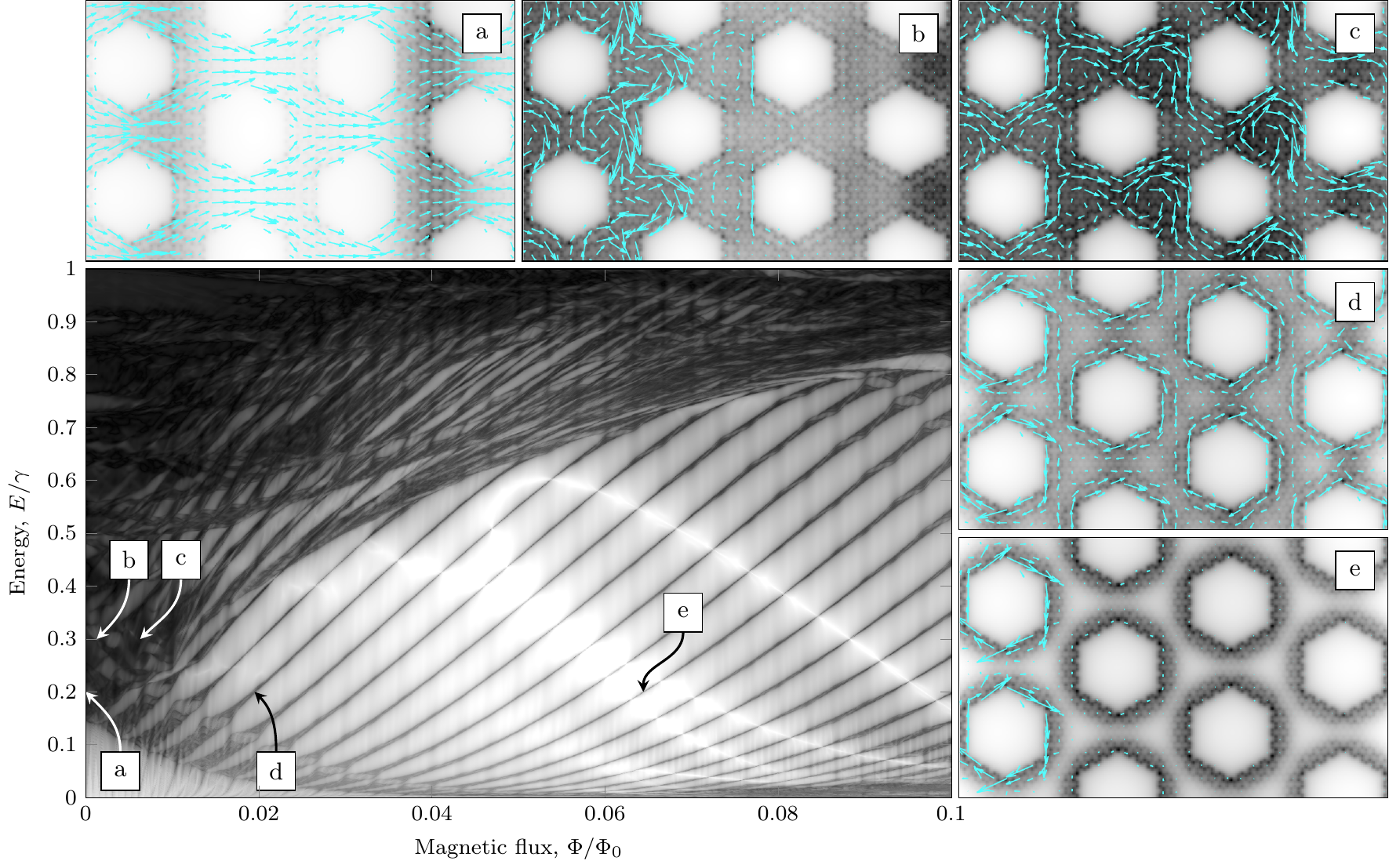}
	\caption{LDOS (gray shading) and bond current (blue arrows) of a \{10,6\} triangular GAB for different $B$-field strengths at energies of $E=0.2\gamma$ (a, d and e) or $E=0.3\gamma$ (b and c). The main panel shows the transmittance of the system. Here, we plot $\sqrt{|log_{10}(T)|}$ in order to enhance the contrast.}
	\label{fig:bondcurrents}
\end{figure*}

Previous transport calculations of GABs without a magnetic field have found their transport gap to be in good agreement with those predicted for Dirac mass barriers (DMBs) \cite{thomsen2014dirac,pedersen2012transport}.
These are modeled using the Dirac approximation with a local mass term in order to open a band gap in the barrier region. 
A derivation of the transmittance of a DMB in a magnetic field is included in the appendix. 
Figure~\ref{fig:k0_trans} shows a comparison between the transmittance of GABs with that of DMBs and gapped graphene with similar gap sizes. 
Note that care must be taken in the $B\rightarrow0$ limit, as the magnetic length then tends to infinity. We note that our $B=0$ T results are consistent with the non-magnetic DMB expression in Ref.~\onlinecite{pedersen2012transport}. 
An excellent qualitative match is seen between the DMB and the gapped graphene barrier in almost all cases.
The match between these simplified models and GABs is quite good near the onset of the gap, particularly for smaller antidots. However, discrepancies appear as the energy is increased towards higher order GAB features, as the antidot size increases, and as the field is increased further (not shown). The DMB and gapped graphene models are therefore good for approximating the transport gap given that the magnetic field is not too large.

\subsection{Comparison with DOS}
Figure~\ref{fig:transmittance_dos_comparison_armchair} shows a comparison between DOS and transmittance of $\{L,6\}$ GABs for four different lattice configurations as well as for a gapped graphene model. $L$ was chosen such that the neck widths were approximately the same ($\simeq 1.3$ nm) for all lattices. The transport calculations were performed with 4 rows of antidots in the transport direction. The figure shows that the transmittance spectra retains most of the features of the DOS for all lattice configurations and for gapped graphene. 
The gapped graphene model shows no transmittance between the band gap and first Landau level. A similar situation arises in the GABs, where we can identify a \emph{geometric band gap} and a \emph{Landau level gap}, which are outlined for the triangular lattice (top panels in Fig.~\ref{fig:transmittance_dos_comparison_armchair}) with dashed red and yellow lines, respectively.  
The differences between the spectra are greatest for small fields. Notice that transport is not fully suppressed in the band gap regions, due to the finite width of the barrier. We observe rather high transmittance in the geometric energy gap regions of the rotated triangular lattice, while the transport gap appears larger than the band gap for the rectangular lattice. Additionally, there is rather high transmittance in the band gap region of the honeycomb lattice, and the secondary band gap is completely invisible in transport. 

A striking similarity between all GAB lattice configurations is the narrow bands in the Landau level gap region. We will demonstrate that these are due to magnetic edge states, i.e., states that are localized on the periphery of the antidots by the magnetic field, as illustrated in Fig.~\ref{fig:magnetic_edge_state}. According to Eq.~\eqref{eq:cyclotron_radius}, the edge states here all have cyclotron radii which are smaller than the antidot radius. The similarity between the panels of the figure demonstrates that the magnetic edge states are robust against lattice configuration. The reason for the relatively high transmittance of these states is that the antidots are close enough to their neighbors that the states couple between antidots. 
Magnetically induced band gap quenching is observed both in the DOS and in transmittance. The quenching seems to be due to magnetic edge states as the magnetic edge state bands begin to form at the quenched band gap. Band gap quenching may therefore disappear if the distance between antidots is increased sufficiently or if a large degree of disorder is introduced.

Since magnetic edge states are localized on the antidot edge, these are of course absent in the gapped graphene model. The gapped graphene model in Fig.~\ref{fig:transmittance_dos_comparison_armchair} has approximately the same band gap as the $\{10,6\}$ triangular GAB. However, at these $B$-field values, there is little resemblance between their transmittance spectra. For instance, in the GAB, the transport gap is quenched by the magnetic field, while the transport gap is retained in the gapped graphene model. It was argued in Ref.~\onlinecite{pedersen2013hofstadter} that band gap quenching occurs when the magnetic length become sufficiently small that the eigenstates do not sample the lattice sufficiently for the band gap to be fully resolved. In gapped graphene, however, the band gap is not introduced by geometrical effects and is therefore retained. 
Another notable difference between the gapped graphene model and the GAB is that practically all transmittance, except for the Landau levels, is suppressed in the gapped graphene model for large magnetic fields, which is not the case for the GAB. The gapped graphene result is consistent with results by De Martino \emph{et al.} \cite{martino2007magnetic}, who showed that Dirac electrons incident on a wide magnetic barrier (i.e, either wide spatial region or large magnetic field) will be totally reflected by the barrier independent of the angle of incidence. 
The GAB result is also consistent with the results by Xu \emph{et al.} \cite{xu2008magnetic} that magnetic barriers in graphene nanoribbons are unable to completely suppress electron transport due to successive reflections on the nanoribbon edge. As noted earlier, GALs can be viewed as a connected network of graphene nanoribbons, so the similarity to the nanoribbon case is expected.

The periodic features in the transmittance of the gapped graphene model are Fabry-P\'erot-type oscillations, which are a result of the additional phase factor that comes from the magnetic field. Additional calculations show that the oscillations double in frequency when the device length is doubled, hence demonstrating the Fabry-P\'erot-type nature of the oscillations. This type of oscillations in transmittance has previously been observed in graphene nanoribbons in a magnetic field \cite{xu2008magnetic}. Additionally, we observe excellent agreement between the gapped graphene model and the predicted Landau levels. 

\subsection{Magnetic edge states}

In order to show that the narrow bands in transmittance are indeed edge states, we show the bond current and LDOS of a $\{10,6\}$ triangular GAB at different magnetic fields and at different energies in Fig.~\ref{fig:bondcurrents}. It is clear that the bond currents at these bands are localized around the antidots, whereas the bond currents elsewhere are not. The shown bond currents are averaged over small area elements, which is why there appears bond currents inside some of the antidots. Additionally, the lengths of the arrows are scaled such that the longest arrow in all plots have the same length. In case of circular current paths or large transverse currents, this can makes it appear as if the current does not propagate through the barrier and therefore make it seem like the the transmittance should be lower than it is. 

According to Eq.~\eqref{eq:magnetic_edge_state}, the oscillation period of the transmittance with respect to $B$-field only depends on the circumference of the antidot. 
This is in agreement with the observation that the energies of the edge state bands are nearly linearly dependent on the $B$-field, thus giving rise to the same oscillation period for all energies. 
\begin{figure}[htb]
	\centering
	\includegraphics[width=\columnwidth]{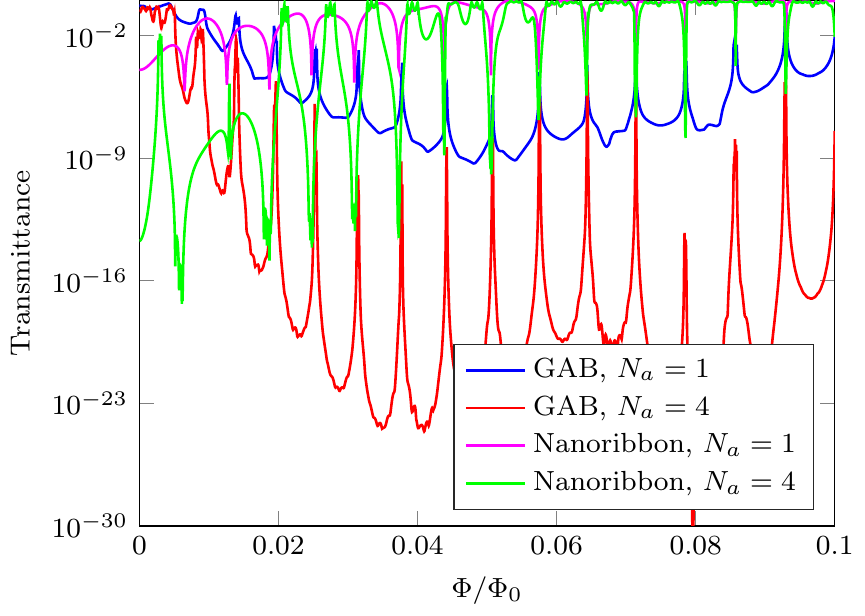}
	\caption{Transmittance as a function of applied magnetic field at an energy of $E=0.2\gamma$ for four different rectangular antidot lattice systems with $N_a$ antidots in the transport direction.}
	\label{fig:magnetic_edge_states_peaks}
\end{figure}
\begin{figure}[htb]
	\centering
	\includegraphics[width=\columnwidth]{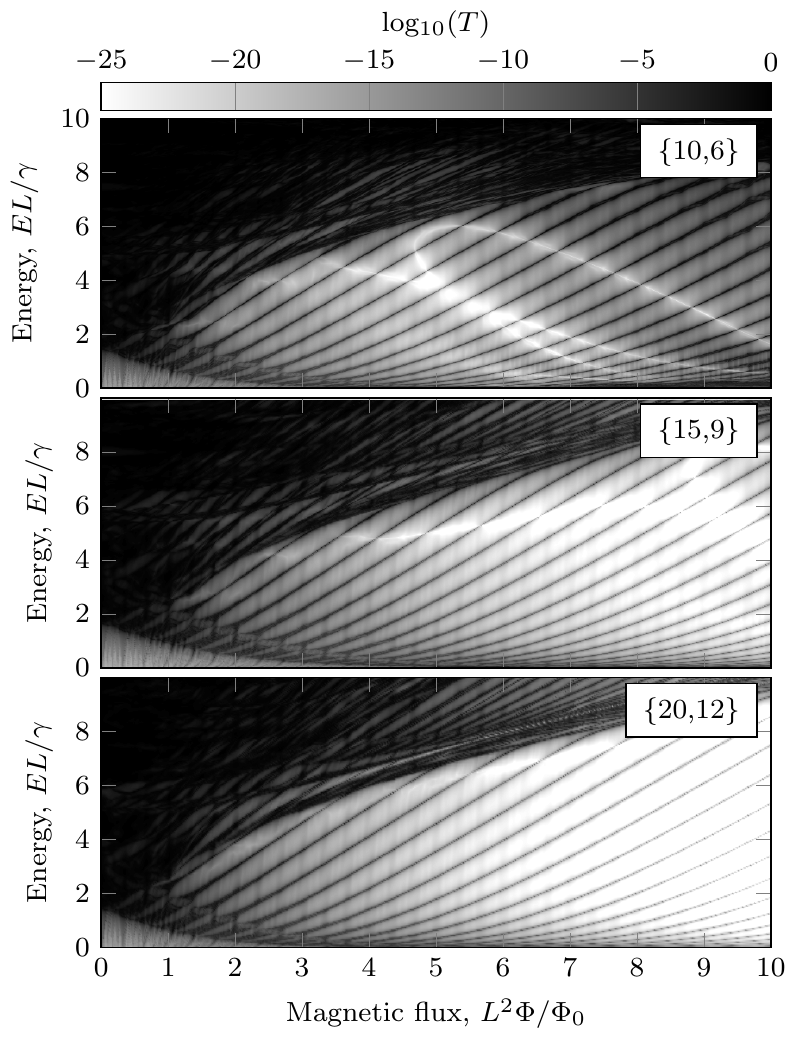}
	\caption{Transmittance of $\{10,6\}$, $\{15,9\}$, and $\{20,12\}$ triangular GABs in scaled units.}
	\label{fig:scaling}
\end{figure}
Increasing the magnetic field corresponds to decreasing the cyclotron radius, which in turn should decrease the average electron distance from the antidot. This is indeed the case, which is apparent when comparing Figs.~\ref{fig:bondcurrents}d and \ref{fig:bondcurrents}e. 
According to Eq.~\eqref{eq:magnetic_edge_state}, the oscillation period is independent of lattice configuration (as confirmed by Fig.~\ref{fig:transmittance_dos_comparison_armchair}), number of antidots and whether the system is periodic or non-periodic, i.e., a graphene nanoribbon. 
In Fig.~\ref{fig:magnetic_edge_states_peaks}, we show the transmittance of GABs and nanoribbons with 1 and 4 rows of antidots in the transport direction. We find indeed that the oscillation period is unaffected by both the number of antidots and periodicity, supporting the validity of Eq.~\eqref{eq:magnetic_edge_state}. For the GABs, we see increased transmittance on the edge state resonances, due to these being the only available states. However, for the nanoribbons, we see decreased transmittance on the edge state resonances. In the nanoribbon case, there is transmission along the edges of the system at these energies without the antidot. Introducing the antidots then gives the electrons a possibility to couple to the antidot magnetic edge states and backscatter. This explains the increased (decreased) transmittance at the edge state resonances for the GAB (nanoribbon) case. Additional calculations show that zigzag antidots with similar circumference have approximately the same oscillation period as armchair antidots (not shown). This demonstrates that the magnetic edge states are additionally robust against antidot chirality.

In Fig.~\ref{fig:scaling}, we compare the transmittance of different $\{L,0.6L\}$ triangular GABs, where the energy and magnetic field axes have been scaled with $L$ and $L^2$, respectively. We see that by plotting on scaled axes, the spectrum is very nearly conserved. 
The scaling with respect to the $B$-field is consistent with Eq.~\eqref{eq:magnetic_edge_state}, which states that the oscillation period due to magnetic edge states is inversely proportional to the square of the circumference. It is remarkable that Eq.~\eqref{eq:magnetic_edge_state} correctly predicts (i) the periodicity of the edge state bands, (ii) the insensitivity to lattice arrangement pf the antidots, and (iii) the behavior under uniform geometry scaling. Additionally, the geometry scaling shows that even though the structures we consider here are probably too small for current experimental realization, our conclusions should hold for larger structures at smaller magnetic fields and energies. Finally, Fig.~\ref{fig:scaling} shows that the transmittance of the magnetic edge states decreases as the distance between antidots is increased, which is expected as these states are localized to the edges of antidots. 

\subsection{Disorder}
\begin{figure}[htb]
	\centering
	\includegraphics[width=\columnwidth]{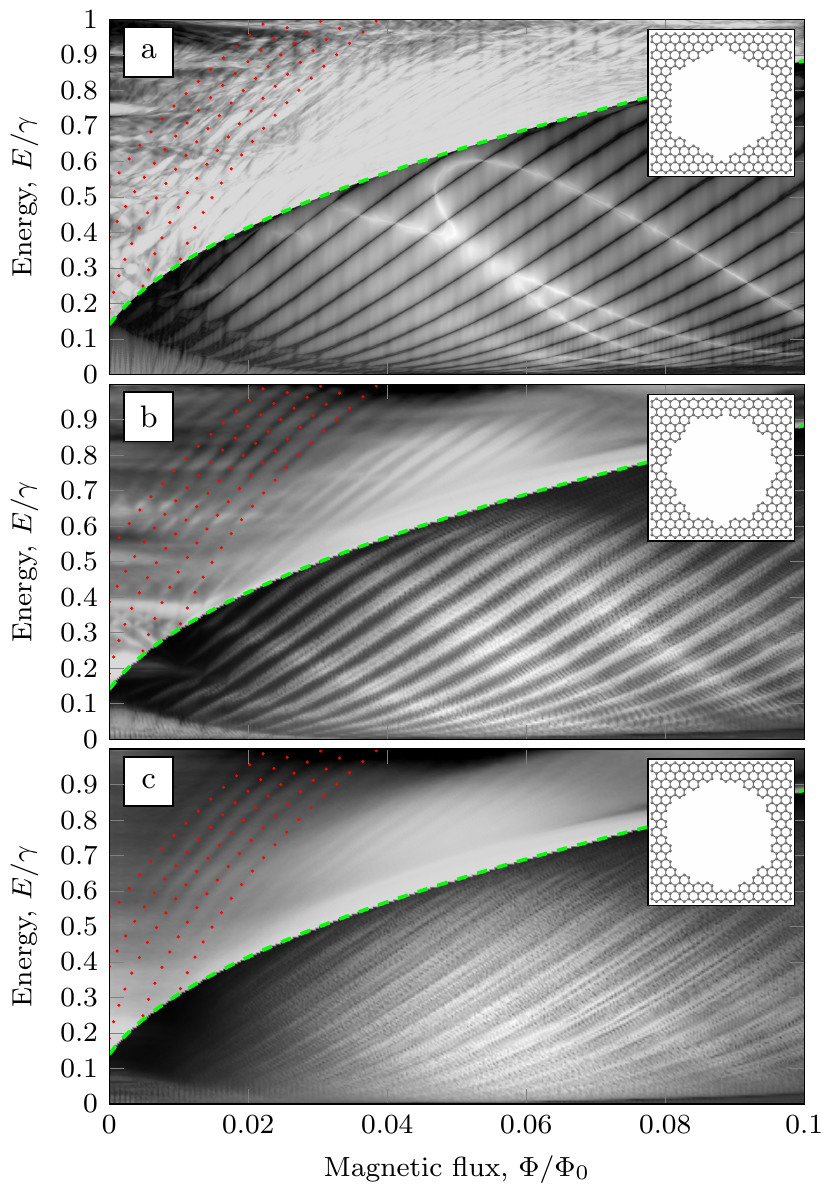}
	\caption{Ensemble averaged transmittance of a) an ordered \{10,6\} triangular GAB and of disordered systems with b) $\sigma=0.5$ and c) $\sigma=1$. The area of the antidots is on average the same in the disordered and ordered systems. An example of the disordered antidots in the two cases is shown as an inset. In order to highlight the features in the plot, we plot $T$ above and $\log T$ below the dashed green line. The dotted red lines are plotted according to $E_i/\gamma = \sqrt{a_i + b\Phi/\Phi_0}$, where $a_i$ and $b\approx 31.668$ were determined by least squares fitting.}
	\label{fig:disorder}
\end{figure}
The systems we have considered until now have been fully ordered. However, experimental samples tend to have varying degrees of disorder. It is therefore important to understand the effects of disorder and find out which features of the transmittance remain. 
The effects of disorder are investigated by ensemble averaging transmittance over different realizations of unit cells with disordered antidots. The antidots were created by first removing six carbon atoms at the locations of the antidots and then iteratively removing edge atoms according to a Gaussian weight profile $w(\vec r) = \frac{1}{N}\sum_i^N e^{-|\vec r - \vec r_i|^2/(2\sigma^2 a^2)}$, where $\vec r$ is the position of the atom, $\vec r_i$ are the centers of the antidots in the ordered system, and $\sigma$ is the standard deviation measured in graphene lattice constants $a$. A large (small) $\sigma$ gives rise to a large (small) degree of disorder. This creates antidots that are roughly centered at the position of the ordered system but with disordered edges. In order to decrease the effects of periodicity, the unit cells are doubled in size in the periodic direction such that there are 8 antidots in the unit cells instead of 4. The ensemble size is determined by convergence testing, and is about 50--100 in the cases we study here.

The ensemble averaged transmittance of two disordered systems with $\sigma=0.5$ and $\sigma=1$, respectively, is shown in Fig.~\ref{fig:disorder} where it is compared to the ordered system. The figure shows that, as the amount of disorder is increased, the rich substructure in transmittance observed in the ordered system is almost completely washed out. However, some of the features of the ordered system do remain. These features form narrow transmittance bands that are highlighted by the fitted red curves in the figure. They are also present in the ordered system, but here they are almost completely disguised by the rich substructure in the transmittance, which is absent in the disordered systems.

Both the Landau levels of pristine graphene, $E_n= \sqrt{2v_F^2\hbar e B n}$\cite{pedersen2013hofstadter}, and the energy levels of a single graphene antidot in a magnetic field\cite{pedersen2012dirac} scale as $\sqrt B$. Therefore, we fit the features in the transmittance spectrum to an expression of the form $E_i/\gamma=\sqrt{a_i+b\Phi/\Phi_0}$, where $a_i$ and $b$ are fitting parameters, which are determined by least squares fitting. In all cases, we find $b\approx 31.668$ although no explanation for this observation has been found. The fitted curves are shown as the dotted red lines on the plots. The fit shows that these features do indeed scale approximately as $\sqrt B$, albeit with an offset.

Both magnetically induced band gap quenching and magnetic edge states in the Landau gap are present for the $\sigma=0.5$ disordered system. However, compared to the ordered system, the initial band gap is decreased and the magnetic edge state bands are broadened. For the $\sigma=1$ disordered system, the edge state bands are broadened sufficiently that they are almost impossible to identify. Additionally, the band gap quenching for this system is less pronounced. The broadening of the magnetic edge state bands is expected as the antidot circumference now differs between individual perforations and, according to Eq.~\ref{eq:magnetic_edge_state}, a variation in circumference of 5~\% will lead to a 10~\% change in the magnetic edge state band position. 
Hence, transmittance features within the Landau gap may be difficult to observe experimentally in disordered samples. In contrast, the robustness of the features above the Landau gap, combined with the long phase-coherence length in graphene, suggests that these states will also be observable in experiments even in the presence of disorder.

\section{Conclusions}

Using a recursive Green's function method, we have calculated electronic transmission and density of states of graphene antidot barriers and graphene antidot lattices, respectively, in magnetic fields. We find, in general, electronic transmission and density of states spectra to be in good agreement. 
We have additionally derived an expression for the transmittance of Dirac mass barriers in magnetic fields and found that this provides a good description of the transport gap of graphene antidot barriers for small antidot sizes and low to moderate field strengths. Calculations of gapped graphene barriers, i.e., graphene with a staggered sublattice potential, are in good agreement with the Dirac mass barrier, and therefore show the same limitations. 

We find that antidots support magnetic edge states, which are robust against variations in lattice configuration, antidot edge chirality, periodicity and number of antidots. Moreover, we observe that these edge states survive a modest degree of disorder. 
The robustness of these states suggests that they will also be observable in experiments even in the presence of disorder. Furthermore, we find that our results scale in a simple manner with system size, thus allowing calculations on small structures to generalize to larger structures.
Additionally, we observe magnetically induced band gap quenching in both density of states and transmittance due to magnetic edge states. In the presence of mild disorder, some fine-structure is washed out but several characteristic and prominent transmission bands are found to survive.

\section{Appendix}
\subsection{Dirac mass barrier}

We can estimate the transmittance through a GAB in a magnetic field by using the Dirac equation with mass term and magnetic field. The mass term and magnetic field are non-zero only in the barrier region, thereby creating a magnetic Dirac mass barrier (DMB).  
We calculate the transmission through this system by matching wave functions at the interfaces on either side of the barrier at $x=0$ and $x=d$. We denote the regions where $x<0$, $0\leq x \geq d$, and $x>d$ as region I, II and III, respectively. 
The wave functions are given by the eigenstates of a generalized Dirac equation, which arises from the substitution $\vec{p}\rightarrow \bm \pi$, where $\bm \pi = \vec{p}+e\vec{A}$ is the generalized momentum
\begin{equation}
\left( \begin{array}{c c} 
\tilde \Delta (x)  & \frac{1}{\hbar} \pi_-^\xi \\ \frac{1}{\hbar} \pi_+^\xi & -\tilde\Delta (x)
\end{array} \right) 
\left( \begin{array}{c} \psi_1 \\ \psi_2 \end{array} \right)
=  k \left( \begin{array}{c} \psi_1 \\ \psi_2 \end{array} \right)\,.
\label{eq:Diraceqn}
\end{equation}
Here, $\tilde\Delta(x) = \Delta (x)/\hbar v_F$, where $\Delta(x)$ is a mass term, which we set equal to $\Delta$ inside the barrier to open a band gap of $2\Delta$, and vanishing elsewhere.
$k=E/\hbar v_F$ is the magnitude of the wave vector corresponding to energy $E$ in graphene in the absence of a $B$-field or mass term. Also, 

\begin{equation}
\begin{aligned}
\pi_\pm^\xi & = \xi \pi_x \pm i  \pi_y
\end{aligned}
\end{equation}
are the standard linear combinations of the $x$ and $y$ components of momenta that occur in the Dirac equation for graphene charge carriers in the $\xi=\pm1$ valley.
From now on we shall assume identical contributions from the valleys and drop the $\xi$ index.
To set a constant magnetic field of strength $B$ in the $\hat{\vec z}$ direction in the barrier, we choose a Landau gauge, see Eq.~\eqref{eq:landau_gauge}. 
Since this gauge, and the system in general, is invariant along $\hat{\vec y}$, we can write the spinor components of the wave function in terms of Bloch functions
\begin{equation}
\left( \begin{array}{c} \psi_1 \\ \psi_2 \end{array} \right) = \left( \begin{array}{c} f(x) \\ g(x) \end{array} \right) e^{i k_y y} \,.
\end{equation}

\paragraph{Region I:} 
As the vector field is zero in region I, the wave functions here are identical to those in pristine graphene. The total wave function can be written as a sum of an incoming (right-going) component of unit amplitude and a reflected (left-going) component, giving
\begin{equation}
\Psi_{I} = \frac{1}{\sqrt{2}} \left[\left( \begin{array}{c} 1 \\ e^{i \theta_k}\end{array} \right) e^{i k_x x} + r \left( \begin{array}{c} 1 \\ -e^{-i \theta_k}\end{array} \right) e^{-i k_x x}\right] e^{ik_yy}\,,
\end{equation}
where $\theta_k = \tan^{-1}(k_y/k_x)$ and $r$ is the reflection coefficient. 

\paragraph{Region II:} 
In region II, the wave functions are solutions of Eq.~\eqref{eq:Diraceqn} with non-zero mass and $B$-field.
Making the substitutions $p_x \rightarrow -i\hbar \partial_x$ and $p_y \rightarrow \hbar k_y$ and rearranging gives
\begin{equation}
\begin{aligned}
\left(- \partial_x^2 + W_+ (x)\right) f(x) &= k^2 f(x) \\
\left(- \partial_x^2 + W_- (x)\right) g(x) &= k^2 g(x),
\end{aligned}
\label{PDE1}
\end{equation}
where
\begin{equation}
W_\pm(x) = \tilde\Delta^2 \pm \frac{1}{l_B^2} + \left(k_y + \frac{x}{l_B^2} \right)^2 \,,
\end{equation}
where $l_B = \sqrt{\hbar/e B}$ is the magnetic length.

By using the substitutions $z = \sqrt{2} \left(k_y l_B + x/l_B\right)$ and $\nu = (k^2 - \tilde\Delta^2)l_B^2/2 - 1$, the expression for $f(x)$ becomes the Weber differential equation
\begin{equation}
\left( \partial_z^2 + \nu + \frac{1}{2} - \frac{z^2}{4} \right) f(x) = 0 \,,
\end{equation}
which has solutions in the form of \emph{parabolic cylinder functions} $D_\nu(\pm z)$. This allows us to write
\begin{equation}
f(x) = \frac{1}{\sqrt{2}} \left( \alpha D_\nu(z) + \beta D_\nu(-z)\right).
\end{equation}
Moreover, $g(x)$ can be related to $f(x)$ using Eq.~\eqref{eq:Diraceqn}, and using the identity $\partial_z D_\nu(z) = \frac{z}{2} D_\nu (z) - D_{\nu+1} (z)$, 
we find
\begin{equation}
g(x) = \frac{i}{l_B(k + \tilde\Delta)} \left[ \alpha D_{\nu+1}(z) - \beta D_{\nu+1}(-z) \right].
\end{equation}
The full wave function in region II is then
\begin{equation}
\Psi_{II} = \frac{1}{\sqrt{2}}\left( \begin{array}{c} \alpha D_\nu(z) + \beta D_\nu(-z)  \\ \frac{\sqrt{2}i}{l_B(k + \tilde\Delta)} \left[ \alpha D_{\nu+1}(z) - \beta D_{\nu+1}(-z) \right] \end{array} \right) e^{ik_yy}\,.
\end{equation}

\paragraph{Region III:} 
In region III, the magnetic field and mass terms are set to zero again. 
However, unlike e.g. Klein tunneling problems where the wave function has a similar form to region I, here we must account for the constant vector potential remaining in this region. The vector potential can not be set to zero in this region, as this would imply an infinite magnetic field in the interface between region II and III. 
We define a wave vector
\begin{equation}
\mathbf{K} = K_x \hat{\vec x} + \left(k_y + \frac{e B }{\hbar}d\right) \hat{\vec y}
\end{equation}
in this region, and enforcing conservation of energy, which is equivalent to conservation of the magnitude of the momentum $K = k$, gives
\begin{equation}
K_x = \sqrt{k_x^2 - \frac{d^2}{l_B^4} - 2 \frac{d}{l_B^2} k_y}\,.
\end{equation}
The wave function in region III is then
\begin{equation}
\Psi_{III} = \frac{t}{\sqrt{2}} \left( \begin{array}{c} 1 \\ e^{i \theta_K} \end{array} \right) e^{i (K_x x + k_yy)} \,.
\end{equation}

\paragraph{Boundary matching:} 
Continuity of the spinor wave function components at the interfaces gives the following set of simultaneous equations which can be solved for $r, \alpha, \beta$ and  $t$
\begin{equation}
\begin{aligned}
1 + r & = \alpha D_\nu(z_0) + \beta D_\nu(-z_0)  \\
t e^{iK_x d} & = \alpha D_\nu(z_d) + \beta D_\nu(-z_d) \\
e^{i \theta_k} - r e^{-i \theta_k} & = \frac{\sqrt{2}i}{l_B(k + \tilde\Delta)} \left( \alpha D_{\nu+1}(z_0) - \beta D_{\nu+1}(-z_0) \right)  \\
t e^{i (\theta_K + K_x d)} & = \frac{\sqrt{2}i}{l_B(k + \tilde\Delta)} \left( \alpha D_{\nu+1}(z_d) - \beta D_{\nu+1}(-z_d)\right).
\end{aligned}
\end{equation}
These four equations are all linear in the coefficients, which makes it straightforward to formulate them as a matrix problem and solve for the coefficients numerically. 
We can then calculate the reflectance and transmittance as $R=|r|^2$ and $T=|t|^2 \text{Re}\{K_x/k_x\}=1-R$. 
The $K_x/k_x$ factor is necessary in order to account for the change in longitudinal momentum. Note that the expressions for $R$ and $T$ are exactly the same as those used in optics.

\section*{Acknowledgments}
The authors would like to thank Mikkel Settnes for valuable discussions. Furthermore, the authors gratefully acknowledge the financial support from the Center for Nanostructured Graphene (Project No. DNRF103) financed by the Danish National Research Foundation and from the \mbox{QUSCOPE} project financed by the Villum Foundation. 
\bibliography{literature}

\end{document}